\documentclass[twocolumn,prd,nofootinbib,superscriptaddress,longbibliography]{revtex4} %

\newcommand\ForInternalReference[1]{}
\newcommand\SkipForEarlyCirculation[1]{}

\newcommand\SkipPP[1]{}
\usepackage[colorlinks=true,citecolor=blue,urlcolor=blue]{hyperref}
\usepackage{verbatim}
\usepackage{graphicx}
\usepackage{dcolumn}
\usepackage{bm}
\usepackage{color}
\usepackage{xcolor}
\usepackage{xspace}
\usepackage{url}
\usepackage{float}
\usepackage{multirow}
\usepackage{adjustbox}
\usepackage[normalem]{ulem}
\usepackage{bm}
\usepackage{placeins}
\usepackage{afterpage}
\usepackage{amsmath,amssymb}
\usepackage{acronym}
\newcommand\optional[1]{}
\acrodef{NR}[NR]{Numerical Relativity}

\usepackage{color}
\definecolor{amber}{rgb}{1.0, 0.75, 0.0}
\definecolor{orange}{rgb}{1.0, 0.5, 0.0}
\definecolor{amaranth}{rgb}{0.9, 0.17, 0.31}

\newcommand{\mc}{{\cal M}}

\def\ltsima{$\; \buildrel < \over \sim \;$}
\def\simlt{\lower.5ex\hbox{\ltsima}}
\def\gtsima{$\; \buildrel > \over \sim \;$}
\def\simgt{\lower.5ex\hbox{\gtsima}}

\newcommand{\UT}{\affiliation{Center for Gravitational Physics, The University of Texas at Austin, Austin, Texas 78712, USA}}

\begin{document}

\title{Adapting a novel framework for rapid inference of massive black hole binaries for LISA}
\author{Aasim Jan}
\UT
\author{Richard O'Shaughnessy}
\affiliation{Center for Computational Relativity and Gravitation, Rochester Institute of Technology, Rochester, New York 14623, USA}
\author{Deirdre Shoemaker}
\UT
\author{Jacob Lange}
\UT
\begin{abstract}
The Laser Interferometer Space Antenna (LISA) is designed to detect a variety of gravitational-wave events, including mergers of massive black hole binaries, stellar-mass black hole inspirals, and extreme mass-ratio inspirals.  LISA’s capability to observe signals with high signal-to-noise ratios raises concerns about waveform accuracy. Additionally, its ability to observe long-duration signals will raise the computational cost of Bayesian inference, making it challenging to use costly and novel models with standard stochastic sampling methods without incorporating likelihood and waveform acceleration techniques. In this work, we present our attempt to tackle these issues. We adapt {\tt RIFT} for LISA to take advantage of its embarrassingly parallel architecture, enabling efficient analysis of large datasets with costly gravitational wave models without relying on likelihood or waveform acceleration. We demonstrate that our open-source code can accurately infer parameters of massive black hole binary signals by carrying out a zero-noise injection recovery using the numerical relativity surrogate model {\tt NRHybSur3dq8}. By utilizing all available $m\neq0$ modes in the inference, we study the impact of higher modes on LISA data analysis. We study the impact of multiple massive black hole binary signals in a dataset on the inference of a single signal, showing that the selected source's inference remains largely unaffected. Furthermore, we analyze the LDC-1A and blind LDC-2A datasets from the Radler and Sangria challenge of the LISA data challenges. When eschewing specialized hardware, we find {\tt NRHybSur3dq8} injection-recovery takes approximately $20$ hours to complete, while the analysis of Sangria and Radler datasets takes about $10$ hours to complete.

\end{abstract}
\maketitle

\section{Introduction}
\label{sec:Intro}

The Laser Interferometer Space Antenna  (LISA) \cite{amaroseoane2017laser} presents exciting scientific opportunities by operating in the millihertz frequency range, enabling the detection of sources currently inaccessible to ground-based observatories like the LIGO-Virgo-KAGRA network \cite{2015CQGra..32g4001L,2015CQGra..32b4001A,2021PTEP.2021eA101A}. Among these sources are massive black hole binaries (MBHBs) \cite{PhysRevD.93.024003}, stellar-mass black hole inspirals \cite{PhysRevLett.116.231102}, extreme mass ratio inspirals (EMRIs) \cite{PhysRevD.66.064005, PhysRevD.75.024005,PhysRevD.95.103012}, and galactic white dwarf binaries (GBs) \cite{refId0}. Ground-based detectors are unable to detect these low-frequency signals due to limitations imposed by factors such as seismic noise \cite{PhysRevD.93.112004} and the restricted size of the detectors, challenges that LISA is designed to overcome.

While LISA's ability to detect a diverse array of signals promises exciting scientific insights, it also introduces challenges not encountered with ground-based detectors. LISA is anticipated to detect a large number of signals \cite{Amaro_Seoane_2023}, with durations ranging from days to years. The combination of long signal durations and a high detection rate will lead to overlapping signals, necessitating simultaneous characterization of gravitating sources through global fitting pipelines \cite{PhysRevD.107.063004,katz2024efficientgpuacceleratedmultisourceglobal,strub2024globalanalysislisadata}. 
Moreover, such large datasets will require efficient analysis algorithms to manage the increased computational cost of Bayesian inference. Furthermore, some of these signals will possess significantly high signal-to-noise ratios (SNR), which will require accurate waveform models \cite{lisaconsortiumwaveformworkinggroup2023waveformmodellinglaserinterferometer} to be used in inference to avoid biases that could influence scientific conclusions. Other challenges include accurately modeling the LISA response for unbiased inference of source parameters.  Moreover, the data is also expected to contain nonstationary and non-Gaussian noise, as well as glitches and gaps, all of which can complicate analysis. Successfully addressing these challenges is essential for maximizing LISA's scientific output.

The primary focus of the LISA mission is the detection of MBHBs, with recent studies indicating a detection rate of $1$ to $20$ per year \cite{PhysRevD.93.024003, PhysRevLett.117.101102,Salcido_2016,Katz_2019,Bonetti_2019}. MBHBs are anticipated to offer valuable insights into a range of astrophysical, cosmological, and fundamental physics questions \cite{Arun_2022}, with SNRs ranging from hundreds to thousands. Although such high SNR signals will be rich in scientific information, the accuracy of the extracted information will be limited by the finite accuracy of waveform models and the numerical relativity waveforms used in their construction.

Given the importance of MBHBs, considerable effort has been dedicated to analyzing them with prior studies utilizing the Fisher matrix approach \cite{PhysRevD.57.7089,PhysRevD.70.042001,PhysRevD.71.084025,PhysRevD.75.089902,PhysRevD.79.104023} to estimate statistical uncertainties in MBHB parameters. However, this approach by construction cannot handle multimodal distributions, is only applicable in high SNR limit, and has been shown to not encompass necessary information to make general statements about parameter estimation of MBHBs \cite{PhysRevD.91.104001,PhysRevD.78.064005,PhysRevD.77.042001}. As such the community has made a collective effort to develop and utilize Bayesian inference tools. The first study to carry out fully Bayesian analyses on inspiral, merger, ringdown MBHB signals (albeit nonspinning MBHBs) 
was \cite{Marsat_2021}. Since then numerous works have analyzed complete inspiral, merger, ringdown MBHB signals using Bayesian methods. In \cite{PhysRevD.101.124008,PhysRevD.105.044055} a complete analysis of the LISA Data Challenge (LDC) dataset, LDC-1A, including the initial search was carried out. The work in \cite{PhysRevD.102.023033} expanded upon the work of \cite{Marsat_2021} to include aligned spins. Different groups presented their solutions to the LDC-2A Sangria dataset by using global fit \cite{PhysRevD.107.063004,katz2024efficientgpuacceleratedmultisourceglobal,strub2024globalanalysislisadata}, template bank search followed by inference \cite{weaving2023adaptingpycbcpipelineinfer} and inference of a single signal \cite{hoy2023bilbyspacebayesianinference}.  The analysis in \cite{hoy2024rapidmultimodalparameterestimation} significantly decreased inference time by making intuitive arguments to constrain the search parameter space consistent with the observed data. 
All these Bayesian analyses used phenomenological models \cite{Ajith_2007,PhysRevD.77.104017,PhysRevLett.106.241101} to generate template waveforms, except \cite{Marsat_2021} which used a frequency domain reduced-order-model of {\tt EOBNRv2HM} \cite{PhysRevD.84.124052}. Also, most of these studies benefited from likelihood acceleration techniques like heterodyning \cite{2010arXiv1007.4820C,zackay2018relativebinningfastlikelihood,PhysRevD.104.104054}. While heterodyning can speed up analysis by a factor of $10^2$ \cite{PhysRevD.104.104054, hoy2023bilbyspacebayesianinference}, it requires knowledge of a high likelihood fiducial point 
and can produce an observable bias in the recovered posterior distributions, even when the true value is used as the fiducial point \cite{hoy2023bilbyspacebayesianinference}. 

While theoretical studies have explored the effects of inaccuracies in waveform models \cite{PhysRevD.76.104018} and numerical relativity waveforms \cite{PhysRevD.104.044037} on MBHB data analysis, there has been limited work assessing this impact using a full Bayesian approach. 
Therefore, in this paper, we adapt {\tt RIFT} \cite{Pankow:2015cra,gwastro-PENR-RIFT,PhysRevD.96.104041,gwastro-PENR-RIFT-GPU,Wofford:2022ykb} a highly parallelizable, grid-based parameter estimation code for LISA, henceforth called {\tt LISA-RIFT} (open-source code repository \cite{RIFT_LISA_git}). {\tt RIFT} is one of the several parameter inference algorithms that have been effectively utilized to infer binary system parameters from gravitational wave (GW) observations \cite{PhysRevX.9.031040,PhysRevX.11.021053,2021arXiv210801045T,2021arXiv211103606T}. Its main advantage lies in its ability to rapidly and accurately infer parameters using costly or novel models, without relying on techniques to accelerate waveform generation \cite{P_rrer_2014} or likelihood evaluation 
\cite{PhysRevD.104.044062,2010arXiv1007.4820C,Antil:2012wf,PhysRevD.87.124005,PhysRevLett.114.071104,zackay2018relativebinningfastlikelihood,PhysRevD.104.123030,PhysRevD.104.104054}, which often serves as a bottleneck for stochastic sampling methods. {\tt RIFT}'s unique approach also facilitates the study of waveform systematics by enabling direct likelihood comparisons, and model selection by generating low-cost, high-accuracy model evidence, making it an ideal tool for extensive waveform systematics studies for LISA.

In this paper, we discuss our modifications to {\tt RIFT} to carry out inference of LISA signals, including the implementation of the full time-frequency dependent LISA instrument response. 
We show that {\tt LISA-RIFT} can accurately infer MBHB parameters by recovering a multimode signal generated using {\tt NRHybSur3dq8} \cite{varma_surrogate_2019} at zero-noise realization. Additionally, we investigate how the presence of multiple MBHB signals in a dataset affects the inference of a single signal.  We also analyze the LDC-1A and LDC-2A datasets from the Radler and Sangria challenge \cite{ldcwebsite,baghi2022lisadatachallenges}, discussing the impact of instrumental and galactic binary foreground noise on MBHB inference.

The paper is organized as follows: in Sec.~\ref{sec:LISA-RIFT}, we briefly outline {\tt RIFT}’s approach to parameter estimation, describe the implemented LISA instrument response, detail improvements in the initial grid generation, and discuss efforts to enhance user-friendliness. In Sec.~\ref{sec:results}, we present the results from analyzing a multimode zero-noise MBHB signal using {\tt NRHybSur3dq8}, our study on the impact of multiple MBHB signals on the inference of a single signal, and the analysis of zero-noise LDC-1A and noisy LDC-2A datasets. In Sec.~\ref{sec:discussion}, we discuss {\tt LISA-RIFT}’s performance. Finally, in Sec.~\ref{sec:conclude} we summarize our main findings and suggest directions for future research.

\section{RIFT and adapting it for LISA}
\label{sec:LISA-RIFT}

In this section, we provide an overview of the algorithm in Sec.~\ref{ssec:overview}, briefly discuss the LISA response in Sec.~\ref{ssec:response}, outline improvements to the initial grid generation in Sec.~\ref{ssec:grid}, and present a diagnostic tool to improve accessibility in Sec.~\ref{ssec:friendly}.

\subsection{Brief overview}
\label{ssec:overview}
{\tt RIFT} \cite{Pankow:2015cra,gwastro-PENR-RIFT,PhysRevD.96.104041,gwastro-PENR-RIFT-GPU,Wofford:2022ykb} is principally a two-stage iterative process. The first stage involves evaluating the marginalized likelihood  $\mathcal{L}_\text{marg}(\bm{\Lambda}_\alpha|d)$ of data $d$ for each point $\bm{\Lambda}_\alpha$ on a proposed grid of binary parameters. This grid is constructed on intrinsic parameters, which for a quasicircular system includes primary mass $m_1$, secondary mass $m_2$, spin of the primary black hole $\bm{a_{1}}$ and spin of the secondary black hole $\bm{a_{2}}$. For LISA analysis, the grid also includes sky location parameters; ecliptic latitude $\beta$ and ecliptic longitude $\lambda$. The remaining parameters $\bm{\theta}$ are integrated to compute $\mathcal{L}_\text{marg}(\bm{\Lambda}_\alpha|d)$. $\bm{\theta}$ includes the luminosity distance $D_L$, polarization angle $\psi$, inclination $\iota$, coalescence phase $\phi_c$, and coalescence time $t_c$. The integration for $\psi, \iota, \phi_c, D_L$, is carried out through Monte Carlo integration, while time is numerically integrated using Simpson's rule within a user-defined window centered around reported coalescence time $t_\text{ref}$. The equation for  $\mathcal{L}_\text{marg}(\bm{\Lambda}_\alpha|d)$ is
\begin{equation}
    \mathcal{L}_\text{marg}(\bm{\Lambda}_\alpha|d) = \int \mathcal{L}(\bm{\Lambda}_\alpha, \bm{\theta}|d) p(\bm{\theta})d\bm{\theta}.
\end{equation}
Here, $p(\bm{\theta})$ represents the priors over $\bm{\theta}$ variables. This integration is made possible by expressing GW signals in terms of modes $h_{\ell m}(m_1,m_2,\bm{a_1},\bm{a_2};t)$, which are associated with the spin-weighted spherical harmonic decomposition of radiation in all possible directions. The plus $h_+(t)$ and the cross polarizations $h_\times(t)$ of the GW are related to $h_{\ell m}(t)$ as:

\begin{equation}
    h_+(t) - ih_\times(t) = \sum_{\ell=2}^{\ell_\text{max}} \sum_{m=-\ell}^{\ell} \frac{D_{\text{ref}}}{D_L}h_{\ell m}(t) Y^{-2}_{\ell m}(\iota,\phi_{\text{c}}).
\end{equation}
In this equation, $D_\text{ref}$ is set to $200$ Gpc. 
Given the highly localized probability mass for $\bm{\theta}$, adaptive importance sampling is utilized in the integration.

In the second stage of {\tt RIFT}'s iterative process, the marginalized likelihood values are interpolated giving the code access to a function that can output marginalized likelihood values at any point in the $\bm{\Lambda}$ parameter space. Using the continuous marginalized likelihood distribution ${\mathcal L}_{\rm marg}(d|\bm{\Lambda})$ and prior $p(\bm{\Lambda})$, the marginalized posterior is constructed via Bayes' theorem:
\begin{equation}
\label{eq:post}
    p_{\rm post}({\bm \Lambda})=\frac{{\cal L}_{\rm marg}(d|\bm{\Lambda} )p(\bm{\Lambda})}{\int d\bm{\Lambda} {\cal L}_{\rm marg}(d|\bm{\Lambda} ) p(\bm{\Lambda} )}.
\end{equation}
The integral in the denominator is computed using Monte Carlo integration. 
Adaptive importance sampling is employed to sample in the $\bm{\Lambda}$ parameter space and to compute the integral. The integral represents the evidence and allows us to compute the Bayes factor, which facilitates the comparison of signal analyses between two models and indicates which model is preferred.

For the subsequent iteration, the grid is generated using a subset of posterior samples from the previous iteration, with an additional expansion of the grid to ensure that regions of high likelihood that might have been missed can be explored.
The iterative process continues until convergence is achieved between two successive iterations, with multiple criteria in place to assess convergence. In the final iteration, the posteriors for the marginalized parameters $\bm{\theta}$ are constructed. For each $\bm{\lambda_\alpha}$, several $\bm{\theta}$ are retained based on their weights. The final posterior for $\bm{\theta}$ is then constructed by combining all of these samples.


\subsection{LISA response}
\label{ssec:response}
LISA  will orbit the sun with a period of one year and during this orbit its orientation will continuously shift relative to the source. 
For long-duration signals, this motion and variation in the orientation will imprint on the signal observed by LISA and needs to be considered in data analysis. 
In contrast, LIGO detects binary black hole signals with durations on the order of seconds, allowing the detector to be treated as static in an inertial frame during detection; and, therefore, during inference when generating template waveforms for comparison with LISA data, a time- and frequency-dependent transformation function must be applied.

In this work, we implement a LISA response model based on the formalism presented in \cite{marsat2018fourierdomain,Marsat_2021}. This response formalism takes into account the time-frequency dependence of the LISA response. It is used to construct single-link observable $y_{slr}$ which represents a laser shift between the transmitting spacecraft $s$ and the receiving spacecraft $r$ along the link $l$. These observables are constructed by applying the transfer function $\tau^{\ell m}_{slr}(f, t_{\ell m})$ to each spherical harmonic mode $h_{\ell m}$ as

\begin{equation}
    \tilde{y}_{slr}(f) = \sum_{\ell,m} \tau^{\ell m}_{slr}(f, t_{\ell m})\tilde{h}_{\ell m}(f).
\end{equation}

The transfer function depends on the parameters $\beta, \lambda, \iota, \phi_{\text{c}}, \psi, t_\text{ref}$. The time-frequency dependence for each harmonic, $t_{\ell m}(f)$, is determined from the stationary-phase approximation by taking the derivative of the phase with respect to frequency:
\begin{equation}
    t_{\ell m}(f) = -\frac{1}{2\pi}\frac{d \Psi_{\ell m}(f)}{df} + t_{\text{ref}}.
\end{equation}
Here, $\Psi_{\ell m}(f)$ is the phase of the frequency domain mode $h_{\ell m}(f)$. These single-link observables are affected by laser noise, which is several orders of magnitude larger than the signal \cite{Tinto_2005}. To address this, time delay interferometry (TDI) is employed, allowing the use of these observables to construct a new set of virtual interferometric observables with significantly suppressed laser frequency noise. 
The LDC datasets provide three such TDI observables: $X,Y,Z$ \cite{PhysRevD.59.102003,Armstrong_1999,PhysRevD.62.042002,Tinto:2000qij,Dhurandhar:2001tct}. However, these observables exhibit correlated noise properties and therefore, in this work, we use the uncorrelated combinations $A,E,$ and $T$. These observables are related to the TDI observables $X,Y,Z$ as follows \cite{Vallisneri:2004bn}:

\begin{equation}
    A = \frac{1}{\sqrt{2}} (Z-X),
\end{equation}
\vspace{-1em}
\begin{equation}
    E = \frac{1}{\sqrt{6}} (X-2Y+Z),
\end{equation}
\vspace{-1em}
\begin{equation}
    T = \frac{1}{\sqrt{3}} (X+Y+Z).
\end{equation}

Although the contribution to the total SNR by the $T$ channel is significantly smaller compared to the $A$ and $E$ channels for the signals analyzed in this work, it is included in our analysis as its contribution becomes significant at higher frequencies \cite{PhysRevD.65.082003}.

We validated our implementation of the response against {\tt BBHx} \cite{PhysRevD.102.023033,PhysRevD.105.044055,michael_katz_2021}. We found our implementation has an excellent agreement with {\tt BBHx}. To quantify this, we computed mismatches between our response and response obtained using {\tt BBHx} and found the mismatches to be of $O(10^{-10})$. Mismatches range between $0$ and $1$, where a mismatch of $0$ indicates that the two series are identical. At this negligible level of mismatch, any bias introduced would only become significant \cite{PhysRevD.78.124020} at an SNR significantly higher than what LISA is expected to observe. For generating GW modes, we utilize the existing {\tt RIFT} waveform interface, allowing the use of all implemented waveform models for data analysis.  In this work, we used the {\tt LALSimulation} \cite{lalsuite} library to generate modes for {\tt NRHybSur3dq8} \cite{varma_surrogate_2019} and {\tt IMRPhenomD} \cite{Khan_2016}.

\subsection{Well placed initial grid}
\label{ssec:grid}
The first step in a run is to set up an initial grid, which requires prior knowledge of a point with significant likelihood 
and the SNR. The choice of this point does not affect the final posterior distributions. Ideally, this information is obtained from a search pipeline, which, for LISA, could also be derived from a global fit pipeline. Using this information, we create a grid on chirp mass $\mc_{c} = (m_1 m_2)^{3/5}/(m_1+m_2)^{1/5}$, symmetric mass ratio $\eta = (m_1 m_2)/(m_1+m_2)^2,~a_{1z},~a_{2z},~\beta$ and $\lambda$. We estimate the bounds of this grid by calculating the Fisher errors using the 2-PN equation from \cite{PhysRevD.52.848}. We also multiply the estimates for each parameter by a factor ($>1$) to address possible systematics and to compensate for the tendency of these equations to underestimate statistical errors for any given signal. For each of our runs, we construct a regular grid on $\mc_{c}$ and $\eta$, while randomly sampling points within the ranges of $a_{1z}, a_{2z}, \beta$ and $\lambda$. The multiplicative factor on each parameter ensures that the grid is compact enough to avoid unnecessary exploration in low-likelihood regions, yet broad enough to enclose the entire posterior distribution and account for potential systematics. 

\subsection{Run analysis: a user-friendly approach}
\label{ssec:friendly}
{\tt RIFT}'s unique approach to parameter estimation can make diagnosing and assessing the health of a run challenging for new users. To make this process more accessible, we have developed a code called {\tt plot\_RIFT.py}. This code generates a variety of plots designed to aid users in diagnosing the status of their {\tt LISA-RIFT} runs effectively. In addition to the visual aids, {\tt plot\_RIFT.py} also produces a comprehensive diagnostics file. This file reviews all relevant metrics, offering solutions if any evaluation does not meet the expected standard. By providing visual diagnostics and detailed reporting, we aim to make the code more accessible to new users.

\section{Results}
\label{sec:results}
\begin{figure}
    \includegraphics[scale=0.33]{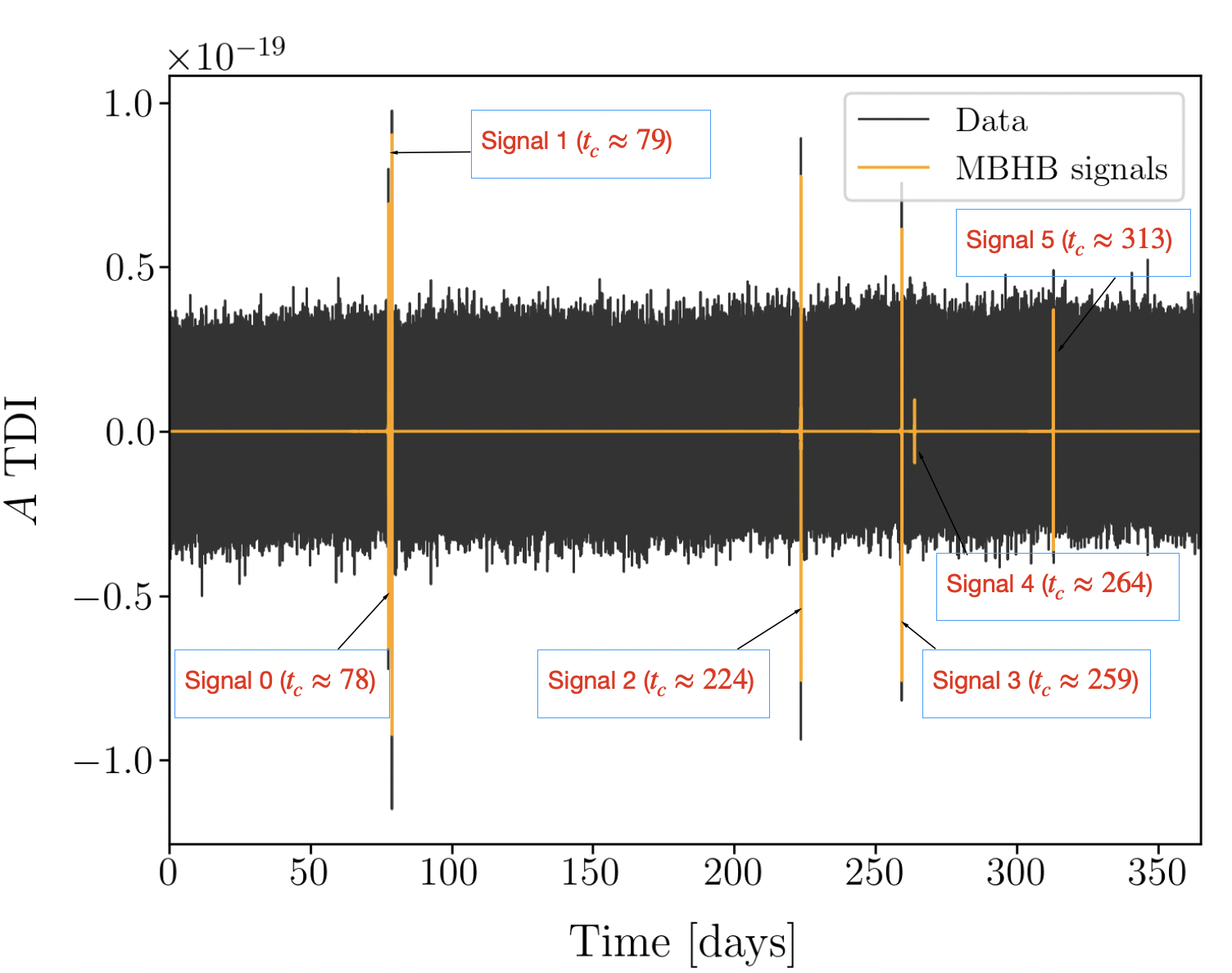}
    \caption{LDC-2A dataset: This figure shows the year-long time series of the $A$ TDI observable from the Sangria LDC-2A dataset.  Black represents the data, which includes six MBHB signals, Gaussian instrumental noise, and waveforms from 30 million GBs. The six underlying MBHB signals are highlighted in orange, with the coalescence time (in days) for each signal indicated. Notably, signal-0 and signal-1 merge nearly a day apart.}
    \label{fig:sangria_dataset}
\end{figure}

In this section, we evaluate the effectiveness of {\tt RIFT}’s iterative algorithm and test our modifications to it by inferring parameters of MBHB signals in mock LISA data. We first analyze an {\tt NRHybSur3dq8}   waveform injected at zero-noise. The injected waveform and the recovery templates include all available $\ell \leq 4$ modes, except $m=0$ modes, and the $(5,5)$ mode, to assess the impact of higher-order modes on MBHB inference. We then analyze the LDC-2A (blind) and LDC-1A datasets from the Sangria and Radler challenge respectively. For our analysis of the LDC-2A dataset, we examine the noisy data that includes galactic binary foreground noise. We also include zero-noise analyses, varying the number of MBHB signals in the dataset to study the impact of the presence of multiple signals in the data when inferring the parameters of a single signal. The six MBHB signals in the dataset, with and without noise, are illustrated in Fig~\ref{fig:sangria_dataset}.  We also include the analysis of the zero-noise LDC-1A dataset.

In all our parameter estimation runs, we maintain a uniform prior over all parameters except $\iota$, for which our prior is uniform in $\cos\iota$. To marginalize over time when evaluating the marginalized likelihood, we integrate over a $600$-second interval centered around $t_\text{ref}$. Following \cite{PhysRevD.105.044055}, we define $t_\text{ref}$ at the frequency at which $f^2 A_{2,2}(f)$ reaches its maximum, where $A_{2,2}(f)$ is the amplitude of $\tilde{h}_{2,2}(f)$. A wide time integration window allows us to account for any time shifts that may occur due to differences in the reference frequency convention used to define $t_\text{ref}$. In all the plots, sky and orientation angles are in the Solar System barycenter (SSB) frame rather than the LISA frame.

We employ the {\tt AV} sampler (based on the {\tt Varaha} sampler \cite{PhysRevD.108.023001,tiwari2024varahapromisingsamplerobtaining}) to carry out adaptive importance sampling to evaluate marginalized likelihood and to construct posteriors. Since likelihood evaluation is not a bottleneck, we calculate the full Whittle likelihood \cite{Moran1951HypothesisTI} over the frequency range of $0.1$ mHz to $0.1$ Hz at the full Fourier bin width. 
However, for our analysis of the LDC-1A dataset we integrate up to $0.05$ Hz, which is the Nyquist frequency of that dataset. We confirm the convergence of the algorithm by ensuring that the Jensen-Shannon divergence (JSD) \cite{JS_test} between the posteriors from the last two iterations is equal to or below $0.001$ bit. JSD quantifies the dissimilarity between two posterior distributions,  where a value of $0$ bit indicates identical distributions and $1$ bit signifies completely different distributions. A JSD of $0.007$ bit is considered to indicate a significant difference in the posteriors \cite{PhysRevX.11.021053}. By ensuring a JSD of less than $0.001$ bit between the last two iterations, we ensure that the interpolation errors are comparable to sampling errors, which are of $O(10^{-4})$ bit \cite{PhysRevD.110.024023}. All the posteriors presented here were generated using $20,000$ samples.

\subsection{Zero-noise injection}

\begin{table}[tb]
\centering
\addtolength{\tabcolsep}{15pt} 
     \begin{tabular}{c c}
     \hline
     Parameter & Value\\ [0.5ex] 
     \hline
     $m_1 $ & $3.375\times10^6 [M_\odot]$ \\
     $m_2 $ & $1.126\times10^6 [M_\odot]$  \\
     $a_{1z} $ & $0.60$ \\
     $a_{2z} $ & $0.25$ \\
     $\lambda $ & $0.628$ \\
     $\beta $ & $0.785$ \\
     $D_L $ & $49.41$ [\text{Gpc}]\\
     $\iota$ & $0.785$ \\
     $\phi_c$ & $0.628$ \\
     $\psi$ & $0.449$ \\
     $t_c $ & $10480000.0$ [\text{s}]\\
     \hline
     \end{tabular}
     \addtolength{\tabcolsep}{-20pt}
     \caption{Parameters of {\tt NRHybSur3dq8}  injection: The masses are in the detector frame, with all angles in radians. The sky and orientation parameters are defined in the SSB frame. This MBHB has a mass ratio of approximately $0.334$.}
    \label{tab:NRHyb}
\end{table}

We inject an MBHB signal, with parameters summarized in Table~\ref{tab:NRHyb}, into approximately four months of data sampled at five-second intervals at a zero-noise realization. The signal was generated using numerical relativity surrogate model {\tt NRHybSur3dq8}, allowing us to incorporate a much larger set of higher-order modes in our analysis compared to previous studies \cite{Marsat_2021,PhysRevD.102.023033}, with the injected signal including content from the following modes: (2,2),(3,3),(4,4),(5,5),(2,1),(3,2),(3,1),(4,3) and (4,2). The inference was performed using the same model and with the same mode content. This was done to eliminate the impact of model waveform systematics and isolate the impact of higher modes on data analysis. Both the injected waveform and recovery templates were generated at a $f_\text{min}$ of $0.06$ mHz. To ensure higher-order modes are excited and significantly impact SNR, we set the mass ratio of the injection to approximately $1/3$, observed at an $\iota$ of $\pi/4$. The distance was chosen such that the signal will be observed at a high SNR of 1000. For likelihood computations, we use the {\tt lisa-data-challenge} \cite{ldc_code} python package to generate the power spectral density (PSD), with the specific noise model used being ``SciRDv1". 

\begin{figure*}
    \includegraphics[scale = 0.31]{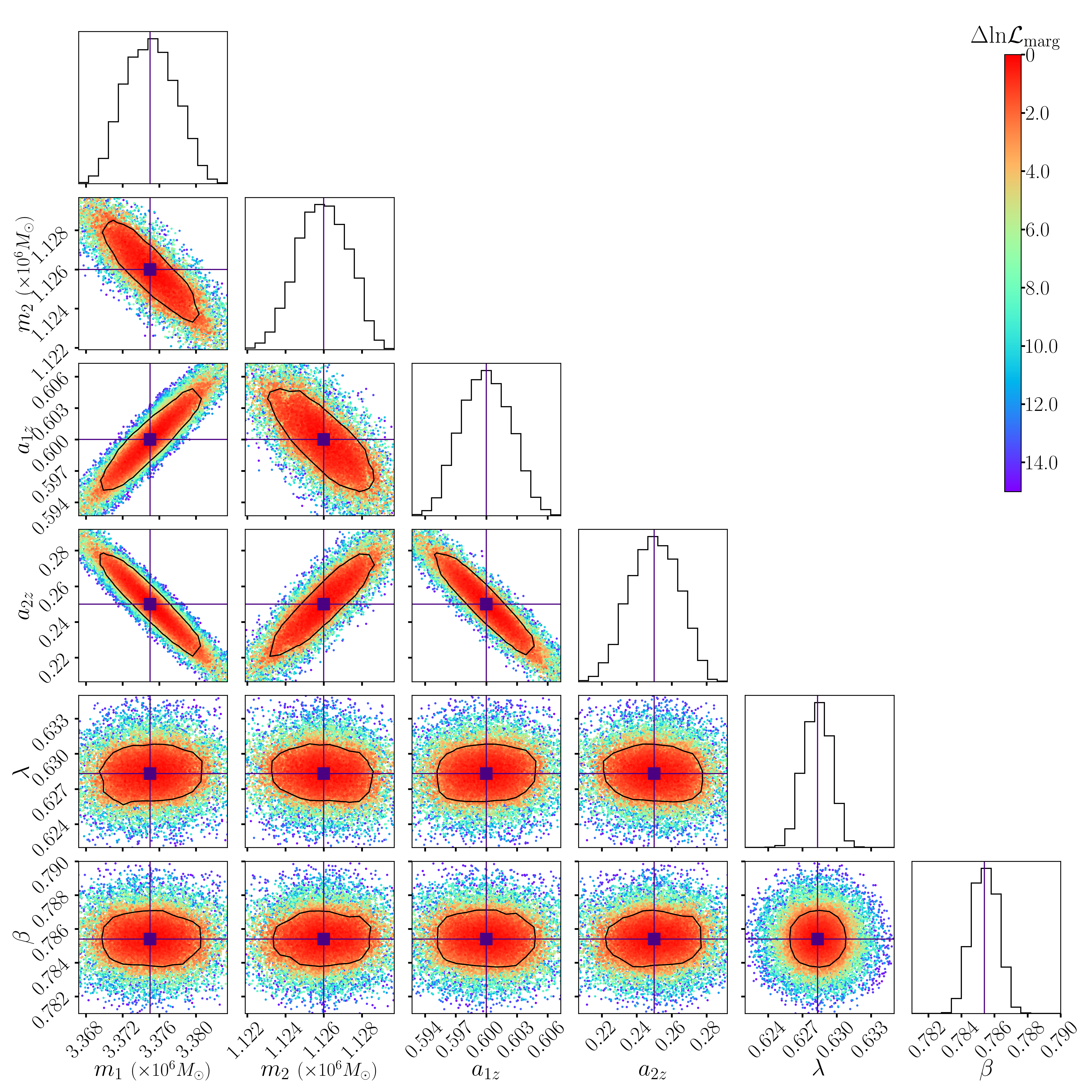}
    \includegraphics[scale = 0.35]{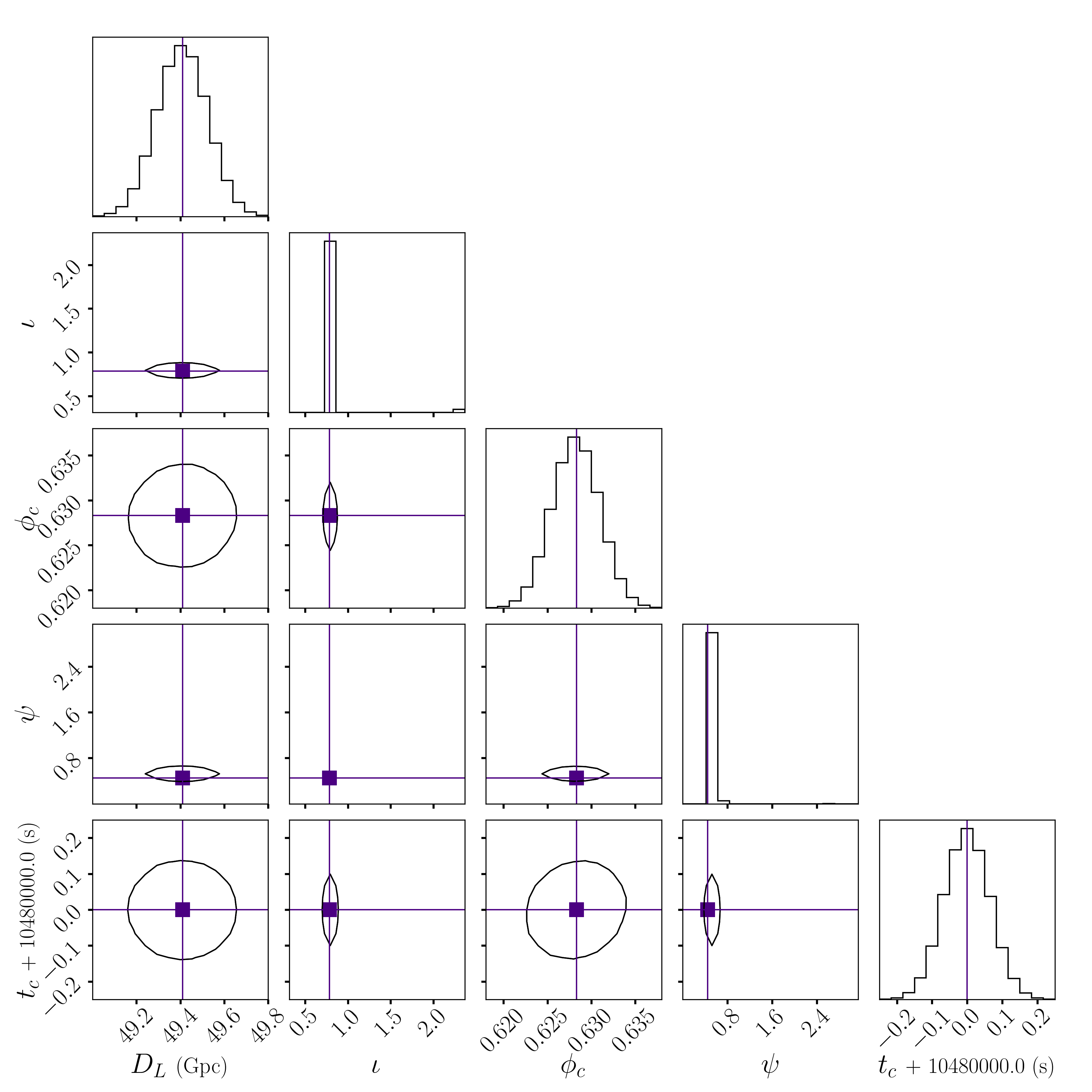}
    \caption{Zero-noise injection recovery using {\tt NRHybSur3dq8}: This figure illustrates the results of our analysis of the four month long zero-noise {\tt NRHybSur3dq8}  injection. For both plots, diagonal panels show the one-dimensional marginal posterior distribution, while contours in the off-diagonal panels show the 90\% credible intervals for the two-dimensional marginal posterior distribution. In the top plot, the colored points represent the $\text{ln}\mathcal{L}_\text{marg}$ values used in constructing the posteriors, within the range of $\Delta\text{ln}\mathcal{L}_\text{marg} \leq 15$ measured with respect to the maximum $\text{ln}\mathcal{L}_\text{marg}$ value. The crosshairs and vertical lines mark the true values. For clarity, the reflected sky mode is not included in this plot.}
    \label{fig:snr_1000_int}
\end{figure*}

The results of our analysis of this mock signal are presented in Fig~\ref{fig:snr_1000_int}. Due to the high SNR and the presence of higher mode content, the posteriors for mass and spins are Gaussian-like and peak at the injected values. The typical bias observed in sky localization for (2,2) only mode analyses \cite{Marsat_2021} is absent, with the posteriors peaking at the true values. Furthermore, when considering the full time and frequency dependence of the LISA response and conducting a complete inspiral-merger-ringdown analysis, only the true and reflected modes\footnote{The reflected sky mode (*) is related to the true sky mode through the following transformations in the LISA ($L$) frame: ~~~~~~~~~$\beta^*_L = -\beta_L$, $\lambda^*_L = \lambda_L$.} for sky position are observed. The reflected mode for this signal, located at $\beta=0.436$ and $\lambda=1.199$, has a low probability; for clarity, we exclude the reflected sky mode from Fig~\ref{fig:snr_1000_int}. Additionally, we note that the degeneracy between $D_L$ and $\iota$ \cite{Usman_2019} is broken, resulting in two Gaussian-like posteriors that peak at the injected values.  In contrast, in Sec.~\ref{ssec:ldc} where the injection and recovery is performed using a (2,2) only mode model, we begin to observe biases in sky location, $D_L$, and $\iota$.

\subsection{LISA data challenge}
\label{ssec:ldc}
We analyze two datasets provided by the LDC working group of the LISA Consortium, specifically from the Sangria and Radler challenges of the LISA data challenges \cite{ldcwebsite,baghi2022lisadatachallenges}. These challenges are designed to foster the development and improvement of analysis algorithms for LISA, with the complexity of the challenges increasing with each new challenge. While these datasets can contain signals from multiple sources, we only focus on the MBHB signals. The MBHB signals in these datasets were produced using the (2,2) only mode aligned spin model {\tt IMRPhenomD}; and, therefore, in our analysis of these two datasets we utilize the same model to generate template waveforms to prevent model waveform systematics from impacting our results.

\begin{table}[tb]
\centering
\addtolength{\tabcolsep}{15pt} 
     \begin{tabular}{c c}
     \hline
     Parameter & Value\\ [0.5ex] 
     \hline
     $m_1 $ & $1.757\times10^6 [M_\odot]$ \\
     $m_2 $ & $1.300\times10^6 [M_\odot]$  \\
     $a_{1z} $ & $0.51$ \\
     $a_{2z} $ & $0.14$ \\
     $\lambda $ & $2.278$ \\
     $\beta $ & $0.293$ \\
     $D_L $ & $33.70$ [\text{Gpc}]\\
     $\iota$ & $1.835$ \\
     $\phi_c$ & $4.228$ \\
     $\psi$ & $1.242$ \\
     $t_c $ & $6712404.5$ [\text{s}]\\
     \hline
     \end{tabular}
     \addtolength{\tabcolsep}{-20pt}
     \caption{Parameters of signal-0 from LDC-2A dataset: The masses are in the detector frame, with all angles in radians. The sky and orientation parameters are defined in the SSB frame. This MBHB has a mass ratio of approximately $0.74$.}
    \label{tab:sangria_param}
\end{table}

\textit{Sangria}: From the Sangria challenge, we analyze the LDC-2A (blind) dataset, which contains six MBHB signals, instrumental noise, and waveforms from GBs; however, the level of instrumental noise and the exact number of GBs are unknown. 
This dataset spans one year and is sampled at five-second intervals, with the $A$ TDI observable plotted in Fig~\ref{fig:sangria_dataset}. Out of the six MBHB signals, we focus on the inference of signal-0, the parameters of which are provided in Table \ref{tab:sangria_param}. 

\begin{figure}
    \includegraphics[scale=0.565]{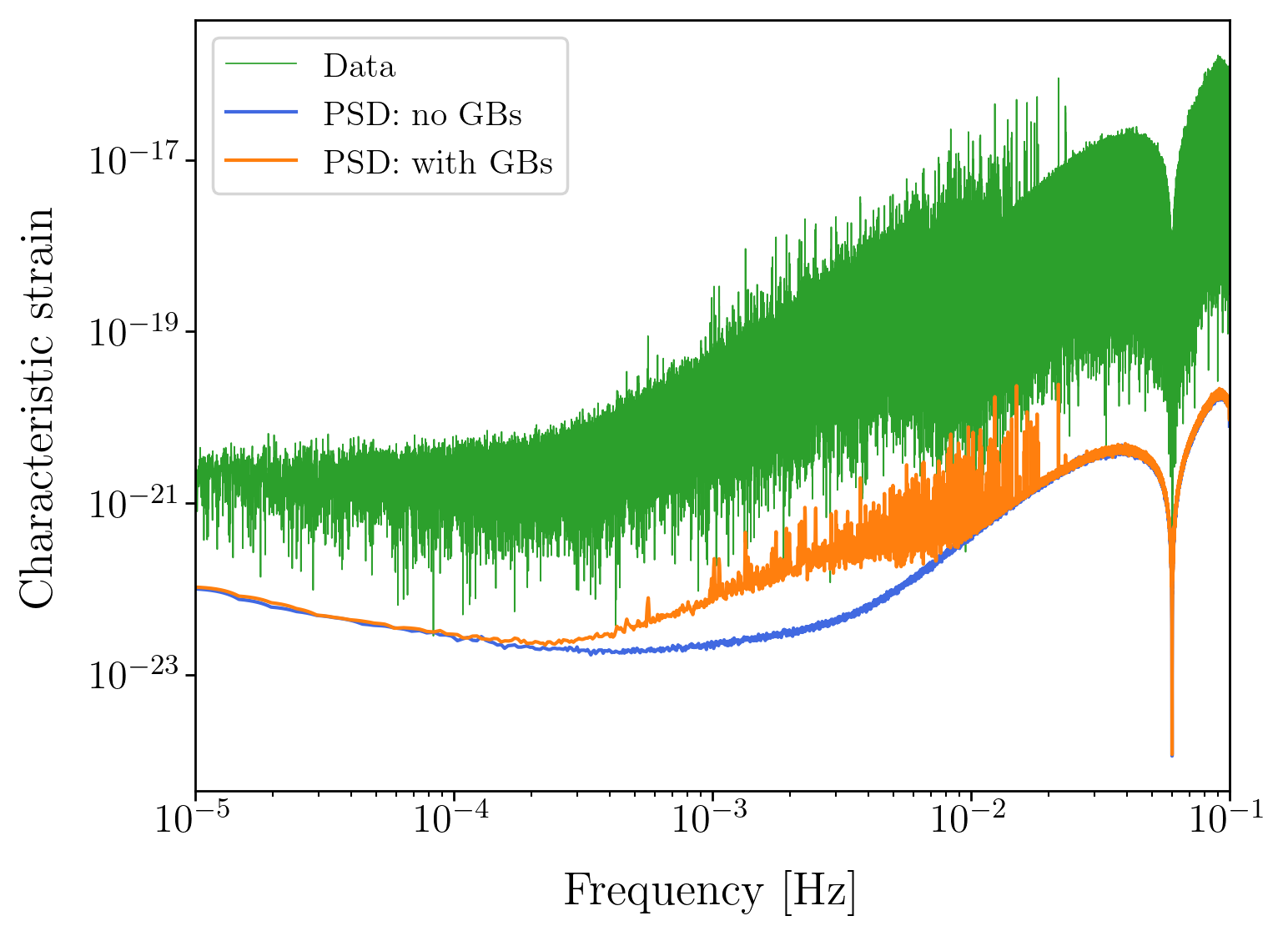}
    \caption{PSDs used in LDC-2A analyses: This figure illustrates the noise amplitudes for two different PSDs (for $A$ channel) used in LDC-2A analyses. One PSD (orange) included GBs in its generation, while the other (blue) did not. The PSD that accounted for GBs was employed for the analysis of signal-0 in the complete noisy dataset, whereas the one without GBs was used for the zero-noise study of signal-0. Additionally, the figure illustrates the characteristic strain of the $A$ TDI observable (green) from the LDC-2A dataset.}
    \label{fig:psd}
\end{figure}

Figure \ref{fig:sangria_dataset} indicates that signal-1 merges one day after the merger of signal-0.  To investigate the impact of nearby MBHB signals on the recovery of signal-0, we first carry out zero-noise inference of signal-0 in three scenarios: in scenario 1 the data only contains signal-0, in scenario 2 the data contains signal-0 and signal-1, and in scenario 3 the data contains all six signals. For this investigation, we generate the PSDs by applying the Welch method to the LDC-2A training dataset, excluding GBs from its generation. The PSD for $A$ channel is plotted in blue in Fig~\ref{fig:psd}, and using these PSDs the SNR of signal-0 is approximately $1277$.

\begin{figure*}
    \includegraphics[scale = 0.31]{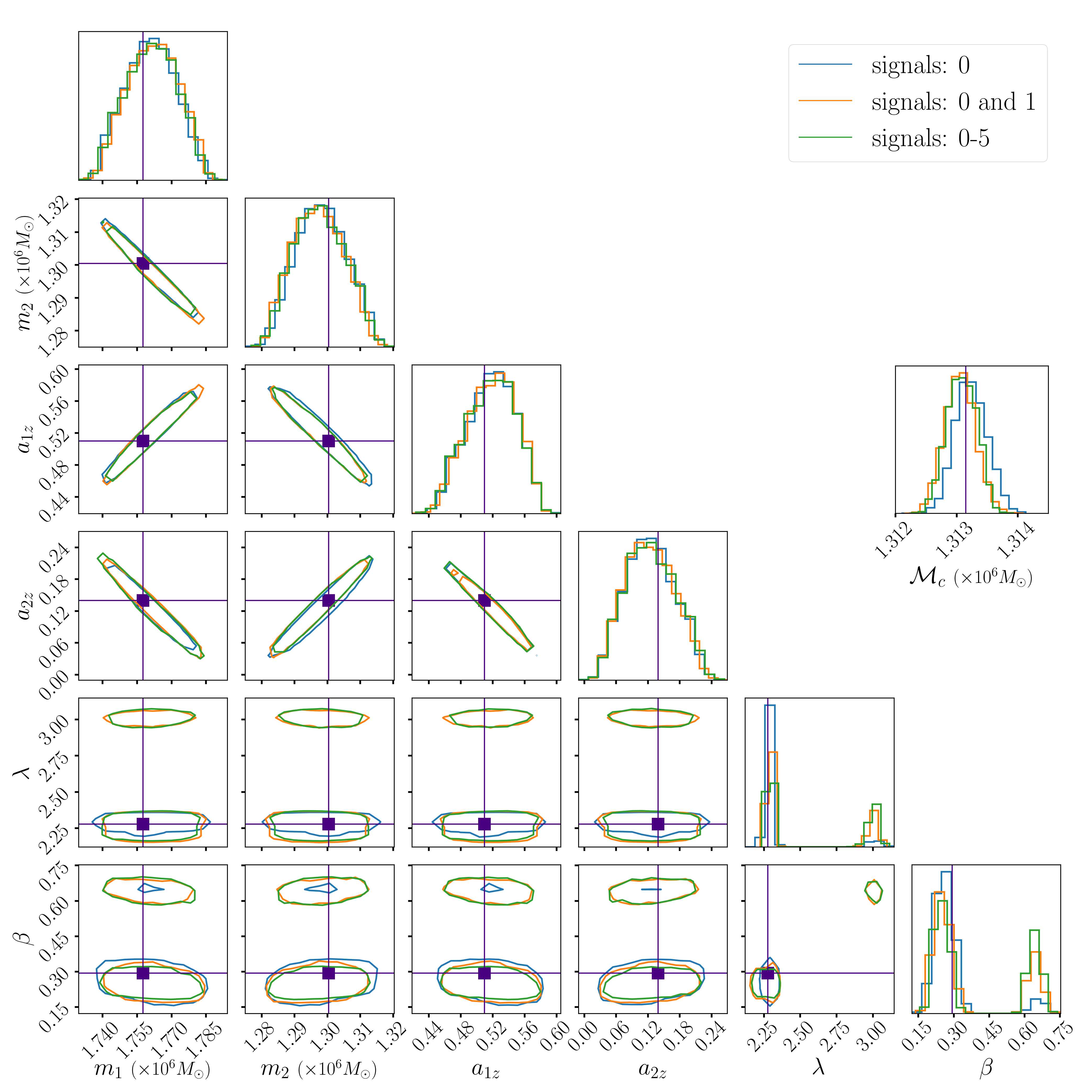}
    \includegraphics[scale = 0.35]{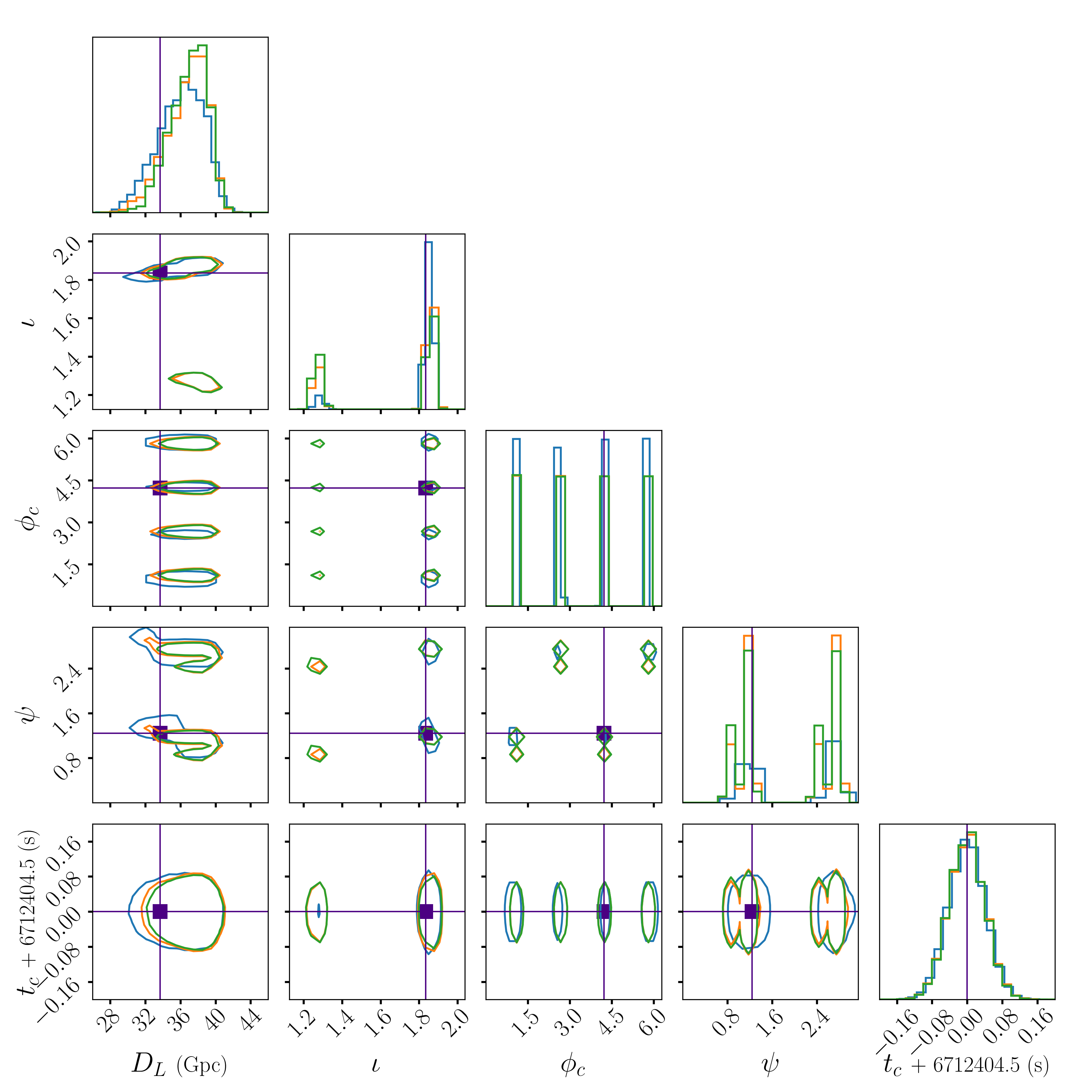}
    \caption{Zero-noise injection recovery of signal-0 in multiple scenarios: This figure illustrates the results of analyzing signal-0 in three scenarios: scenario 1 only has signal-0 in the data (blue), scenario 2 has signal-0 and signal-1 in the data (orange), and scenario 3 has all six signals in the data (green).  For both plots, diagonal panels show the one-dimensional marginal posterior distribution, while contours in the off-diagonal panels show the 90\% credible intervals for the two-dimensional marginal posterior distribution. The crosshairs and vertical lines mark the true values. In all three scenarios, the dataset is one year long. We find that the presence of other signals does not significantly affect the analysis of signal-0.}
    \label{fig:sangra_no_noise}
\end{figure*}

The results of this investigation are illustrated in 
Fig~\ref{fig:sangra_no_noise}. From the figure, we can see that the introduction of other signals has minimal impact on the recovery of the $m_1,m_2,a_{1z}$ and $a_{2z}$. However, the impact is more pronounced if we express the masses as $\mc_c$. To quantify the differences in the posteriors from scenarios 2 and 3 when compared to posteriors from scenario 1, we calculate absolute normalized bias. It is defined as $|\Delta \tilde{\kappa}|/\sigma$ where $\Delta \tilde{\kappa}$ is the difference in the median of $\kappa$ posterior distributions and $\sigma$ is the standard deviation of $\kappa$ posterior from scenario 1. For both scenarios 2 and 3, the normalized bias in $m_1,m_2,a_{1z}$ and $a_{2z}$ is much lower than $0.15$. In contrast, for $\mc_c$, the bias is around $0.55$.
 While a bias greater than $1$ is typically seen as significant, this value is notably larger than those obtained for the fundamental mass and spin parameters. The bias in $D_L$ is approximately $0.35$ for both scenarios. 
 
 Since the sky location posteriors are bimodal, we instead calculate the probability mass in the reflected mode. We find the probability mass in the reflected mode increases from $0.07$ in scenario 1 to $0.28$ in scenario $2$ and to $0.38$ in scenario $3$. While there is a significant increase in the probability contained in the reflected sky mode, the inference still favors the true sky mode. We observe a similar behavior for $\iota$; and, therefore infer that while the inclusion of additional MBHB signals does not significantly alter the inference of signal-0, it does have a measurable impact.

We now analyze signal-0 in the noisy LDC-2A dataset. The PSDs used in this analysis included GBs in their generation and the PSD for $A$ channel is plotted in orange in Fig~\ref{fig:psd}. Using these PSDs, the SNR of signal-0 is approximately $269$. 

\begin{figure*}
    \includegraphics[scale = 0.31]{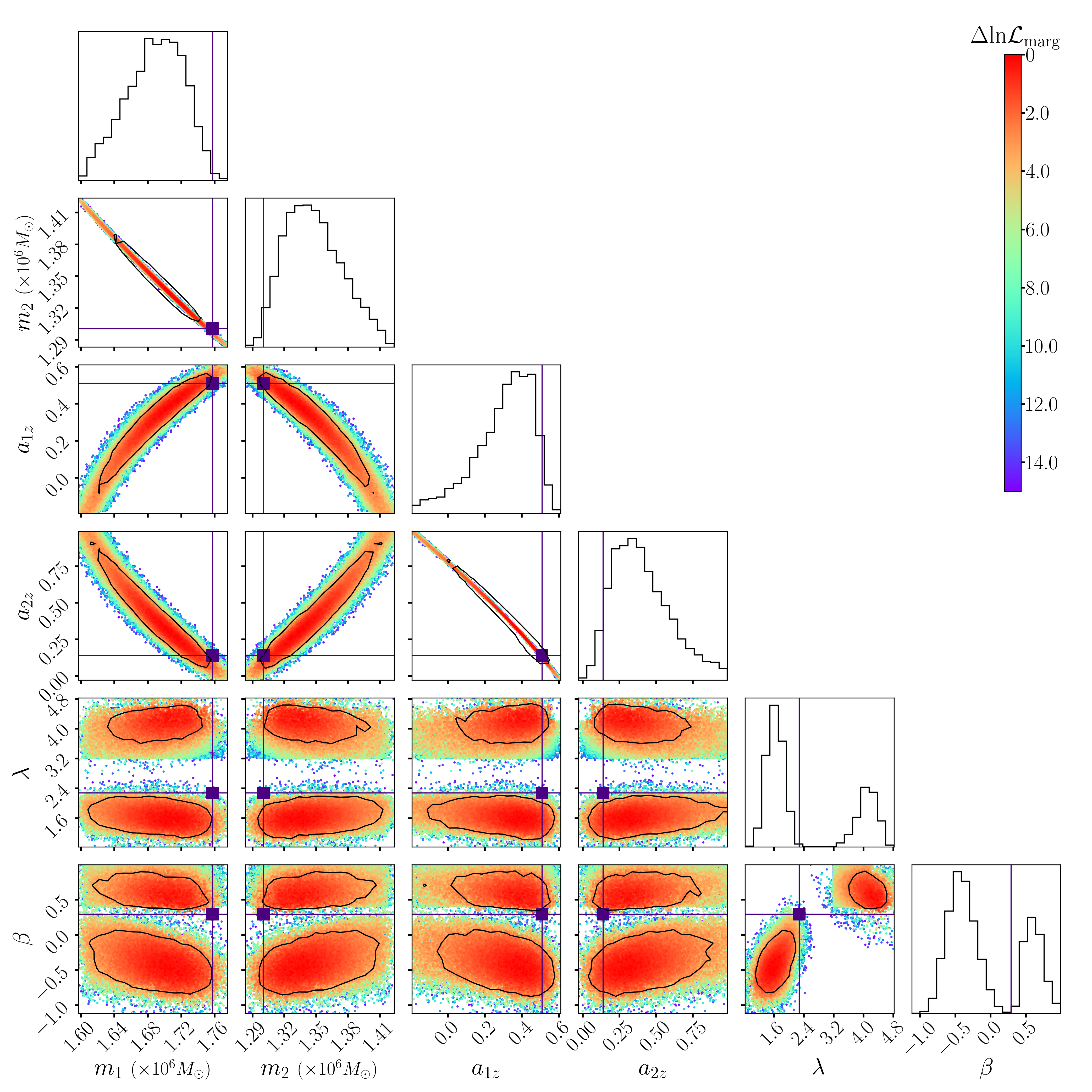}
    \includegraphics[scale = 0.35]{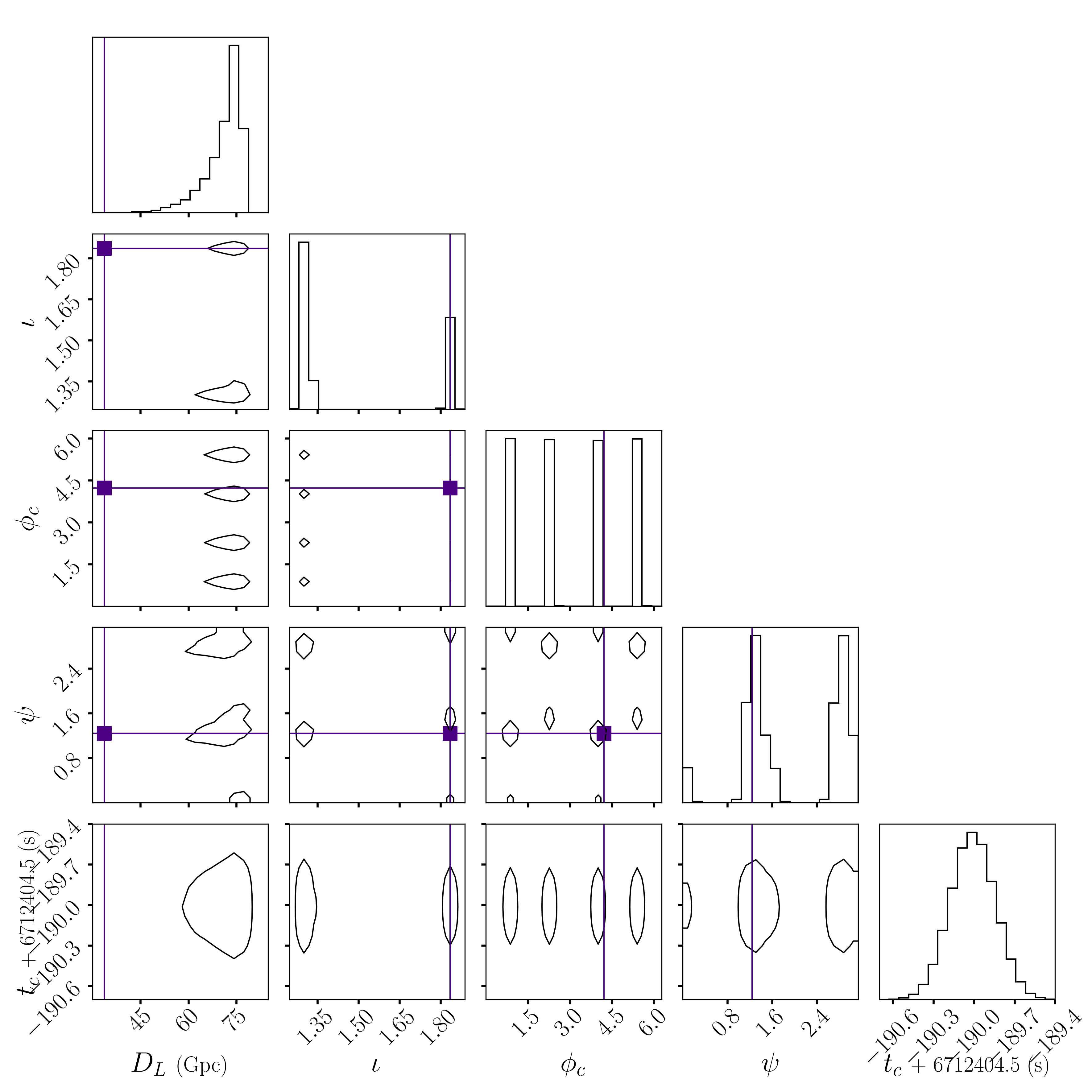}
    \caption{LDC-2A (Sangria) analysis: This figure illustrates the results of our analysis of signal-0 in the year long LDC-2A dataset which contains six MBHB signals, instrumental noise, and GBs.   For both plots, diagonal panels show the one-dimensional marginal posterior distribution, while contours in the off-diagonal panels show the 90\% credible intervals for the two-dimensional marginal posterior distribution. In the top plot, the colored points represent the $\text{ln}\mathcal{L}_\text{marg}$ values used in constructing the posteriors, within the range of $\Delta\text{ln}\mathcal{L}_\text{marg} \leq 15$ measured with respect to the maximum $\text{ln}\mathcal{L}_\text{marg}$ value.  The crosshairs and vertical lines mark the true values. We find that the presence of GBs and noise significantly impacts the inference of signal-0 by shifting the posteriors away from the true value.}
    \label{fig:sangra_noise}
\end{figure*}

The results of our analysis are illustrated in Fig~\ref{fig:sangra_noise}, showing that the true mass and spin values are captured by their respective posterior distributions; with the true values located almost at the edge of the $90\%$ credible interval. Additionally, while the true sky location values are captured by their posterior distributions, they fall outside the $90\%$ credible interval. The parameter that seems to be most impacted by the presence of GBs and noise is $D_L$, as the posterior distribution completely misses the true value. 
 From Fig~\ref{fig:sangra_no_noise}, however, we can see that even in the zero-noise case the $D_L$ posterior distribution is biased due to the degeneracy between $D_L$ and $\iota$, and the introduction of noise and GBs exacerbates this bias. We anticipate the inclusion of higher-order modes will improve $D_L$ recovery in noisy data.  The time posterior also does not include the true value but that is due to the difference in the reference frequency convention used by {\tt LISA-RIFT} and LDC to define $t_\text{ref}$. Consistent with the findings of \cite{weaving2023adaptingpycbcpipelineinfer}, we observe that GBs and noise considerably impact inference, with posterior distributions peaking significantly away from the true values. One way to mitigate this would be to perform a global fit on the data to identify and remove GBs.

\begin{figure}
    \includegraphics[scale=0.35]{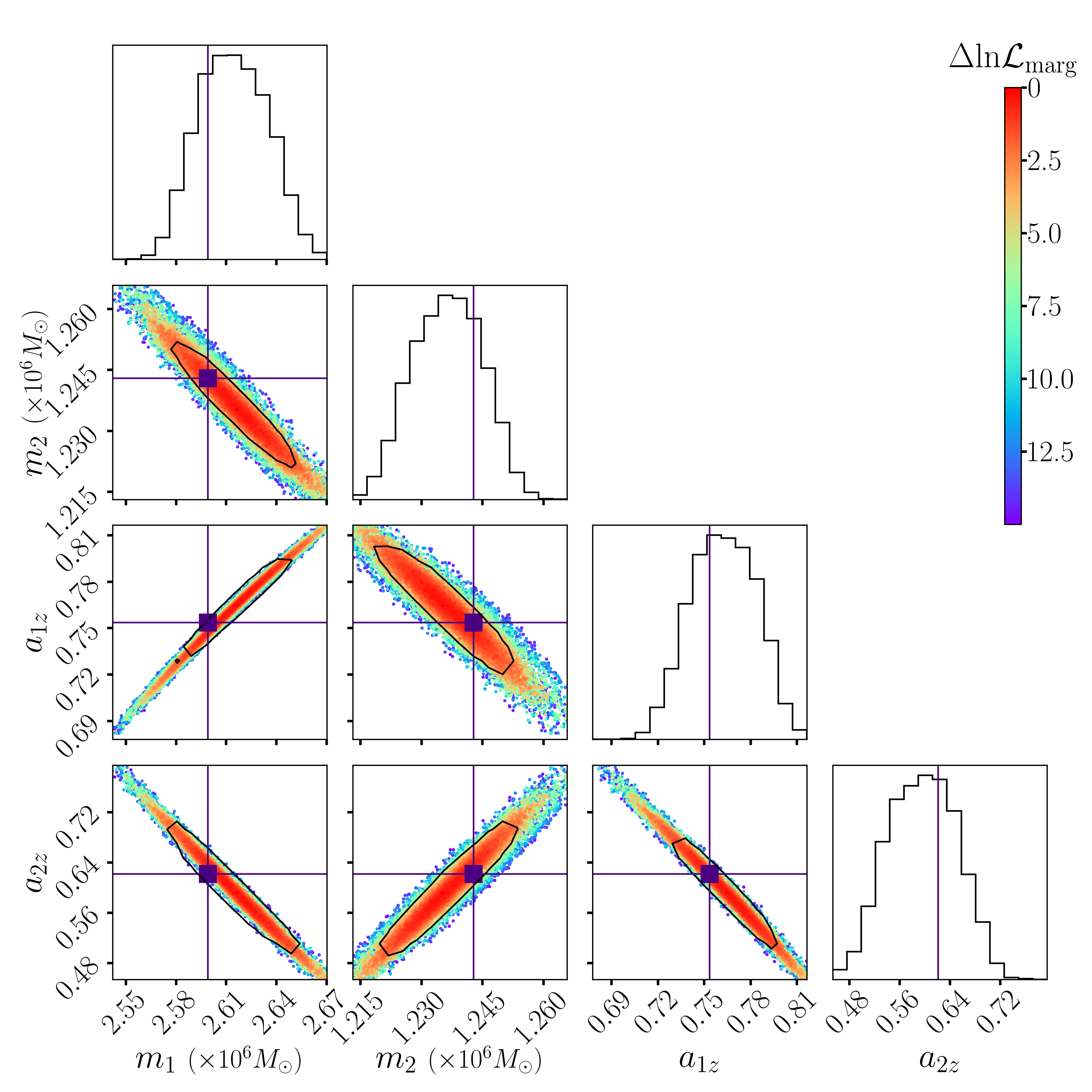}
    \caption{LDC-1A (Radler) analysis: This figure illustrates the results of our analysis of the 1.33 years long LDC-1A zero-noise dataset.  Diagonal panels show the one-dimensional marginal posterior distribution, while contours in the off-diagonal panels show the 90\% credible intervals for the two-dimensional marginal posterior distribution. The colored points represent the $\text{ln}\mathcal{L}_\text{marg}$ values used in constructing the posteriors, within the range of $\Delta\text{ln}\mathcal{L}_\text{marg} \leq 15$ measured with respect to the maximum $\text{ln}\mathcal{L}_\text{marg}$ value. The crosshairs and vertical lines mark the true values.}
    \label{fig:radler}
\end{figure}

\textit{Radler}: From the Radler challenge, we analyze the LDC-1A dataset which includes a single MBHB signal with mass and spin parameters listed in Table~\ref{tab:radler_param}. This dataset spans nearly 1.33 years and is sampled at ten-second intervals.  We analyze the MBHB signal at a zero-noise realization and use ``SciRDv1" PSDs in our likelihood computations. With these PSDs, we find the SNR of the MBHB signal to be approximately $390$.
Figure \ref{fig:radler} illustrates the results of our analysis, indicating that the true values are well within the $90\%$ credible interval.  We observe that our analysis constrains the mass and spin parameters more tightly than \cite{PhysRevD.105.044055}. Additionally, there is a slight shift in the peak for mass posteriors relative to the true values. Further investigation revealed that the $\mc_c$ posterior peaks at the true value, but $\eta$ posterior was slightly shifted, leading to corresponding shifts in the mass posteriors. The color gradient suggests that this shift 
stems from the underlying marginalized likelihood values. This could potentially be caused by differences in data conditioning and/or due to the differences in the likelihood method used; \cite{PhysRevD.105.044055}  used heterodyned likelihood, while our analysis utilized full Whittle likelihood.

\begin{table}[tb]
\centering
\addtolength{\tabcolsep}{15pt} 
     \begin{tabular}{c c}
     \hline
     Parameter & Value\\ [0.5ex] 
     \hline
     $m_1 $ & $2.599\times10^6 [M_\odot]$ \\
     $m_2 $ & $1.243\times10^6 [M_\odot]$  \\
     $a_{1z} $ & $0.753$ \\
     $a_{2z} $ & $0.622$ \\
     \hline
     \end{tabular}
     \addtolength{\tabcolsep}{-20pt}
     \caption{Parameters of the MBHB signal from LDC-1A dataset: The masses are in the detector frame. This MBHB has a mass ratio of approximately $0.478$.}
    \label{tab:radler_param}
\end{table}

\section{Discussion}
\label{sec:discussion}
LISA will detect signals from a variety of sources, and accurately inferring the properties of these sources is vital for maximizing the scientific output of LISA. MBHB signals observed by LISA are expected to have high SNR,  with dataset durations ranging from days to years. While these high SNR signals will be rich in scientific information, analyzing them presents unique challenges. One major challenge is the need for accurate waveform models to be used as templates during inference to avoid introducing biases in signal interpretation \cite{lisaconsortiumwaveformworkinggroup2023waveformmodellinglaserinterferometer}.
Additionally, long datasets significantly increase the computational cost of likelihood evaluations, a key component of any Bayesian algorithm. 

Motivated by the need for a code capable of analyzing long-duration, high SNR signals using models that might be prohibitive to use with standard sampling techniques, we modify the {\tt RIFT} parameter estimation algorithm for LISA. This modification, called {\tt LISA-RIFT}, takes advantage of  {\tt RIFT}'s embarrassingly parallel architecture, enabling the use of costly waveform models at full Fourier bin resolution. We find {\tt LISA-RIFT}'s computational performance to be highly efficient in terms of both wall clock and core time.
For instance, the {\tt IMRPhenomD} analysis shown in Fig~\ref{fig:sangra_no_noise} and Fig~\ref{fig:sangra_noise} only took around $10$ hours of wall clock time and approximately $350$ core-days ($\approx 10^4$ core-hours) of CPU time. The bulk of the computational cost came from around $10^5$ marginalized likelihood evaluations performed during
this analysis, each taking about $30$ seconds. This analysis was performed on a year-long dataset at full Fourier bin resolution. 
In comparison, standard sampling techniques using heterodyned likelihood when applied to LDC datasets typically take anywhere from a few hours \cite{PhysRevD.101.124008} to nearly half a day \cite{weaving2023adaptingpycbcpipelineinfer} of wall clock time to complete their analysis. Considering the $10^2$ speedup provided by heterodyning, these methods would require anywhere from a few days to a month to complete when utilizing the full Whittle likelihood.  Additionally, our approach could be further accelerated by leveraging {\tt RIFT}'s native hardware acceleration \cite{gwastro-PENR-RIFT-GPU}. A similar performance was observed in our analysis using {\tt NRHybSur3dq8}. That analysis, performed on a four month long dataset with both injection and recovery involving nine different GW modes, took around 20 hours of wall clock time.

By significantly decreasing the wall clock time for analyses, we expect {\tt LISA-RIFT} will allow for extensive parameter estimation studies for LISA. Furthermore, it integrates seamlessly with the existing {\tt RIFT} infrastructure, supporting the use of all available MBHB waveform models and allowing for inference with numerical relativity waveforms \cite{PhysRevD.96.104041}. The addition of new waveform models is straightforward.

\section{Conclusions}
\label{sec:conclude}
In this work, we introduced {\tt LISA-RIFT}, a modification of {\tt RIFT} parameter estimation algorithm for rapid and accurate Bayesian inference of MBHB signals. One key modification was the implementation of a full time-frequency dependent LISA response model to enable realistic inference of mock LISA signals. We validated our implementation of the response model against {\tt BBHx}, a publicly available LISA response code, achieving mismatches lower than $10^{-10}$. 

We performed zero-noise inference of an MBHB signal using a highly accurate numerical relativity surrogate model, {\tt NRHybSur3dq8}, which we used for both signal and template waveform generation. We successfully recovered the injected parameters with the recovered posterior distributions peaking at the true values. Further, by utilizing all available $m\neq0$ modes, we demonstrated how higher-order modes can help resolve well-known parameter degeneracies typically encountered in analyses using only the (2,2) mode. 

We  investigated the impact of multiple MBHB signals in the LDC-2A (blind) dataset on the inference of a single signal, signal-0. Our results showed that the presence of additional signals had a minimal impact, with absolute normalized bias in mass and spin posterior distributions remaining below 1.0, even at a high SNR of $1277$. Although the probability mass of the reflected sky mode increased with the addition of other signals, the inference still predominantly favored the true sky mode.

We also analyzed the noisy LDC-2A dataset from the Sangria challenge and found that the presence of GB foreground noise and instrumental noise impacted the inference of signal-0, shifting the posterior peak away from the true value. This reinforces the need for a global fitting procedure to identify and remove GBs from the dataset before analyzing MBHB signals. We also analyzed the zero-noise LDC-1A dataset from the Radler challenge and found that the true values were well within the $90\%$ credible interval. By conducting inference on MBHB signals in both zero-noise and full-noise conditions, we demonstrated the ability to rapidly recover MBHB parameters, completing full inference of an MBHB signal in LDC-1A and LDC-2A datasets in a matter of hours.

In the future, we intend to employ {\tt RIFT}'s native hardware-accelerated options and optimize marginalized likelihood evaluations to further decrease inference time. 
As the lead developers advance the core infrastructure, including the development of faster samplers, we aim for our modifications to work seamlessly with these advancements.

\begin{acknowledgements}
We would like to thank Connor Weaving for his help with the LDC-2A dataset. We thank Sylvian Marsat for addressing our questions regarding the LISA response model, and Michael Katz for making his {\tt BBHx} code publicly available, which was used to validate our implementation of the response model. We thank the anonymous reviewer for their helpful comments. This research was done using services provided by the OSG Consortium \cite{osg07,osg09,https://doi.org/10.21231/906p-4d78,https://doi.org/10.21231/0kvz-ve57} which is supported by the National Science Foundation awards No. 2030508 and No. 1836650. Parts of this code were tested on LLO and LHO cluster of the LIGO scientific collaboration
supported by NSF Grants No. PHY-0757058 and No. PHY-0823459. A.J. and D.S. acknowledge support from NASA Grant No. 80NSSC24K0437. A.J., D.S., and J.L. acknowledge support from NSF Grants No. PHY-2207780 and No. PHY-2114581. 
R.O.S.  gratefully acknowledges support from NSF awards No. NSF PHY-1912632, No. PHY-2012057, No. PHY-2309172, No. AST-2206321, and the Simons Foundation. The work was done by members of the Weinberg Institute and has an identifier of UT-WI-36-2024.

\end{acknowledgements}

\pagebreak

\bibliography{references}

\end{document}